\documentstyle{article}  
\begin{document}  
\title{\bf Twisted Homotopy \\ A Group Theoretic Approach  
}  
\author {  
{\bf   
M.MEKHFI} \thanks {On sabbatical leave from Institut de Physique Univ Es-Senia Oran-  
ALGERIE}\\  
\normalsize Laboratoire De Physique Theorique Es-Senia 31100 Oran ALGERIE \\ And \\International 
Centre For Theoretical Physics      
.Trieste ,ITALY        
}        
\date{}        
\maketitle        
\vskip 4truecm        
\begin{abstract}       
After summarising the physical approach leading to twisted homotopy and after developing the       
cohomological approach further with respect to our previous work we propose a third alternative approach       
to twisted homotopy based on group theoretic considerations.In this approach the       
 fundamental group $\Pi (m) $ isomorphic to Z which describes homotopic        
loops on the punctured plane$ R^2/(0) $ is enhanced  in a special way to the continuous SO(2) group .        
This is performed by letting the parameter of the group  $ m \rightarrow  \lambda       
$       
while keeping its generator unchanged .It is shown that such non-trivial        
procedure has the effect of introducing well defined self-interactions        
among loops which are  at the basis of twisted homotopy where the  angle $ \lambda  $  plays the role of    
the       
self coupling constant.       
.\footnote {\bf KEYWORDS: HOMOTOPY, GROUP THEORY,QUANTUM MECHANICS  \\   
MSC:55Q35; PACS:02.20.Fh ; 03.65.Fd                            }          
\end{abstract}        
\newpage        
\section{Physical approach}       
The punctured plane $ R^2/(0) $ is the  space of interest here        
with its simple  but not trivial topology .From the homotopy \cite{kn:homotopy}\cite {kn:balachandran}       
point of view,the objects which        
describe the        
punctured plane are loops encircling the hole.These loops have a        
structure characterised by the fundamental group we denote $\Pi (n);n \epsilon Z  $ which        
is isomorphic to Z and the winding number W which indicates how much times        
a loop encircles the hole.       
It  will         
help to regard homotopic loops as intrinsic " physical " objects rather         
than just  appropriate mathematical objects used to tag the puncture of the         
plane.Adopting this point of view  \cite{kn:mekhfi1} we will therefore associate to the loop         
of winding number n  the winding number eigenstate          
$ \mid n \; \rangle $. The group elements in turn become operators         
which map the state  $ \mid n \; \rangle $ into the state $ \mid \; n +m \; \rangle $ .         
We therefore summarise the homotopy information on the punctured plane by the set of relations.        
\begin {eqnarray}        
                      W \mid n \rangle &=& n \mid  n  \rangle \nonumber  \\        
                 \Pi ( m ) \mid n \rangle &=& \mid n+m \rangle  \nonumber \\         
\sum_{n\epsilon Z } \mid n \rangle \langle n \mid &=& 1    \\  
\label{eq:homo}          
                                                          \langle n \mid m \rangle&=& \delta_{nm}  \nonumber        
\end {eqnarray}        
\noindent         
The above system describe the standard homotopy of the punctured plane.The states  we have are       
objects        
of definite winding number and as  observable  we have the winding number  W .The group       
elements $ \Pi        
(m )$  play the role of creation and annihilations operators  and may serve   to       
build up        
other type of observables.At this point we may conceive another  alternative to probe the puncture of the 
plane which we    
call twisted homotopy      
which from        
the homotopic point of view is equivalent the above one. In previous  works we have introduced       
twisted homotopy  via  essentially two different ways.The first one \cite{kn:mekhfi1} adopts       
directly the        
point of view that these loops are "physical" in the sense that they may undergo interactions.As   a        
consequence of the assumed interactions new states show up and serve to  probe the punctured       
plane in a        
different but equivalent way.The second one  \cite{kn:mekhfi2} regards the twisted homotopy       
which in        
the case of the punctured plane is isomorphic to twisted homology , as the homological dual to the       
twisted        
cohomology first introduced by E.Witten \cite {kn:witten} in the context of topological quantum      
mechanics.In this   paper        
 we would like to develop to some length the second approach as it was only sketched       
in ~\cite {kn:mekhfi2} and to principally introduce   a third procedure  to derive  twisted       
homotopy  which is based on group theoretical        
considerations.Let us first briefly summarise the first  method.In the physical approach       
an external field is provided to couple to loops through the winding        
charges  with strength $ \beta $ .The coupling to W  has been shown to be  linear in order for the        
field not to destroy the group structure of loops .Thus the hamiltonian is       
\begin {equation}       
H_0 =\beta W       
\label{eq:h0}       
\end {equation}       
\noindent       
In addition self-       
interactions are introduced asking for them to reflect the group        
composition law in the same way as for the external field. It is important to note that the above        
requirement for the interactions to preserve the group structure is the guide line to find the       
appropriate        
interactions similar in this respect to  gauge symmetries in quantum field theories.The hamiltonian  which        
satisfies the above requirements is        
worked out to be of the form        
\begin {eqnarray}       
H( \beta,\lambda ,\rho) &=&\beta W(\lambda ,\rho )\nonumber \\       
W(\lambda,\rho ) &=&W+ \lambda \sum_{m \epsilon Z} \rho ( m) \Pi (m )       
\label{eq:effective}       
\end {eqnarray}       
\noindent       
Where  $ \lambda  \rho ( m ) $  is m dependant self -coupling constant.       
The eigenstates of the above hamiltonian $ \mid  n,\lambda,\rho >$ are related to the unperturbed   one        
through the formula .       
\begin {eqnarray}       
\mid n,\lambda,\rho \rangle &=&e^{-i\lambda J } \mid  n \rangle \nonumber \\       
                 With \;\;\; J&=&-i\sum_{m \epsilon Z/(0) } \frac{ \rho ( m ) }{m}  \Pi ( m )        
\label{eq:newbasis}       
\end {eqnarray}       
\noindent       
This is a unitary transformation provided the "spectral" function  $ \rho( m ) $ obey the       
requirement        
relation for the hamiltonian in eq ~\ref{eq:effective} to be hermetian that is $ \rho ^*( m ) =\rho       
(-m ) $.       
In this new framework the homotopy of the punctured plane is described by a new  basis which       
is        
the unitary transform of the old one  and which define what we called twisted homotopy .The       
latter is        
specified as follows.       
\begin {eqnarray}       
                  J\mid n,\lambda ,\rho \rangle &=& -i\partial _\lambda \mid n, \lambda ,\rho \rangle       
\nonumber        
\\       
 \Pi ( m ) \mid n,\lambda,\rho \rangle &=& \mid n+m ,\lambda,\rho \rangle  \nonumber \\         
\sum_{n\epsilon Z } \mid n,\lambda,\rho \rangle \langle n,\lambda,\rho \mid        
                                                          &=& 1    \\   
    \label{eq:twisted homo}        
    \langle n,\lambda,\rho \mid m,\lambda ,\rho \rangle       
                                                           &=& \delta_{nm}  \nonumber        
\end {eqnarray}        
\noindent         
The states $\mid n,\lambda,\rho > $ which now figure out the "loops " around the hole are no longer       
of        
definite winding number .In   twisted homotopy the winding number is no longer an observable.It       
is the        
generator J built out of the creation and annihilation operator $ \Pi ( m)  $ which is the observable       
and        
which play a role analogous to the   winding number ,where the canonical variables are $ \theta $ and $    
\lambda $ rather than  n  and $ \theta $   \cite{kn:mekhfi2}. \\      
Before passing to the second approach let us mention the possible applications we have found so far of       
twisted homotopy  apart from the already known applications of it in the form of twisted cohomology to    
Morse theory \cite {kn:witten}.In a recent investigation of Bessel functions  in the context of homotopy    
and their      
relations to the punctured plane ~\cite{kn:mekhfi3}we have demonstrated the following correspondence.      
\begin {equation}       
 \mid n \rangle \Leftrightarrow \frac{ J_{n} ( z )}{z^n}      
\label{eq:correspondence1}       
\end {equation}       
\noindent       
Where $ J_n ( z ) $ is the Bessel function  of integer order n.On the other hand the states defining twisted      
homotopy $ \mid n+\lambda \rangle $ written for the case where the spectral function $ \rho ( m )$ is  of    
the form $  \rho ( m  )=( - 1 )^m $      
have been shown to correspond to  reduced Bessel function of real      
orders .That is .      
\begin {equation}       
 \mid n+\lambda \rangle  \Leftrightarrow \frac{ J_{n+\lambda} ( z )}{z^{n+\lambda}}      
\label{eq:correspondence2}       
\end {equation}       
\noindent       
And as a consequence of these correspondences ,reduced Bessel functions become unified through the      
relation ~\ref{eq:newbasis} in a single formula independently  of their orders being integers or reals.      
\begin {equation}       
 \frac{J_{n+\lambda}( z )}{z^{n+\lambda}}=e^{-i\lambda J }\frac{J_n( z )}{z^n}      
\label{eq:unifyingformula}       
\end {equation}       
\noindent      
The second application concerns the observable J which plays a role  analogous to  the winding number    
.In studying      
 observables in cohomological quantum mechanics  defined on the punctured plane we have investigated 
in details in \cite      
{kn:mekhfi2} it turns out that according to the form of the prepotential one has ,the observable  of the      
theory may appear in the form of W which is the winding number operator in standard homotopy or in the      
form of J which is its  analogue in twisted homotopy.      
\section{Cohomological approach}      
The second approach to twisted homotopy starts from the twisted cohomology.Let d be the       
exterior        
derivative which define the De Rham cohomology of the target manifold  M  which will correspond      
to the punctured plane in our case  and define a "twisted" exterior derivative as  \cite {kn:witten}      
\begin {equation}       
d_\lambda [V] = e^{-\lambda V}d e^{\lambda V }       
\label{eq:twisted}       
\end {equation}       
\noindent       
Where $V : M  \rightarrow  R $ is  a Morse function called the prepotential in the framework of     
topological       
(supersymmetric)        
quantum mechanics. At this point we modify the above formula by taking the scaling factor $ \lambda $ 
pure imaginary.That is we consider. 
\begin {equation}       
d_\lambda [V] = e^{-i\lambda V}d e^{i\lambda V }       
\label{eq:twisted2}       
\end {equation}       
\noindent       
The factor i is essential as it will  ensure the hermiticity of the effective winding number which results 
from the twisting.Now we add an important imput to eq ~\ref{eq:twisted2} which is the right form to give 
to  the prepotential V. 
 In the above  cited paper \cite{kn:mekhfi2} we have shown that the most       
general        
prepotential for the punctured plane one can have consistent with the topology is .       
\begin {equation}       
V=K \theta +\Phi ( \theta )       
\label{eq:prepotential}       
\end {equation}       
\noindent       
Where $ \theta $ is the polar angle , K an arbitrary constant  and $ \Phi (\theta ) $ is an arbitrary but        
periodic function of the argument.More importantly the function should be an even function  of        
$ \theta $        
.Note that $ \theta $ is not a periodic function.       
Now on the subspace of  periodic functions defined on the punctured plane  $ u (\theta )$  , the exterior        
derivative takes the form $ d=d\theta \partial_\theta$ .Inserting the form of the prepotential into the       
twisted        
exterior derivative which we may rewrite  as $ d_\lambda[V] =\partial_\theta^\lambda       
[ V ]        
d\theta $ and dividing by $ d\theta$ we obtain       
\begin {equation}       
 \partial_\theta^\lambda [V]=e^{-i\lambda \phi}  \partial_\theta e^{i\lambda \phi}+ i\lambda K       
\label{eq:prepotential2}       
\end {equation}       
\noindent       
To make the transition with the hamiltonian form in eq  ~\ref{eq:effective} we define the        
states        
$\mid \theta > $ which are the Fourier transform of the states $ \mid n > $.       
\begin {eqnarray}    
            u ( \theta ) &=&\langle \theta \mid u \rangle \nonumber \\       
\mid \theta \rangle &=& \sum _ {n\epsilon Z } e^{in\theta} \mid n \rangle  
\label{eq:fourier1}            
\end {eqnarray}       
\noindent       
The winding number operator W acts on the wavefunctions $ u ( \theta ) $ or the states $ \mid \theta     
\rangle $ as $ W=-i\partial _\theta $ While the group elements       
$ \Pi (m ) $  act as $ \Pi ( m ) =e^{-im\theta }$.The function $       
\Phi (        
\theta )$ being periodic  by definition,we decompose it  as a  Fourier series .       
\begin {equation}       
\Phi ( \theta )=\sum _{m \epsilon Z/( 0 ) }\sigma ( m )  \frac { e^{-im\theta }   }{m}      
\label{eq:fourier2}       
\end {equation}       
\noindent       
Where $ \sigma ( m ) $ is odd function of m as a consequence of the evenness of $ \Phi$.We may       
formally        
rewrite $ \Phi ( \theta )$ as the mean value on the theta states  of an operator J as follows.       
\begin {eqnarray}       
\Phi ( \theta ) = \langle \theta \mid J \mid \theta \rangle \nonumber \\       
 J=\sum_{m \epsilon Z/(0) }\sigma ( m )    \frac{ \Pi ( m ) }{m}        
\label{eq:jdefinition}        
\end {eqnarray}         
\noindent         
Multiply both sides of the equation eq ~\ref{eq:prepotential2} by the factor -i, we may arrange the        
resulting equation in operatorial forms where the operators now act on the states $ \mid n \rangle $  .       
\begin {eqnarray}   
 \label{eq:tc} 
\langle \theta \mid W ( \lambda ,\rho )     
         \mid u \rangle&=&\partial ^\lambda_\theta [V] u ( \theta ) \nonumber \\     
W ( \lambda ,\rho ) &=& e^{-i\lambda J} W e^{i\lambda J} + \lambda K  \\  
&=&W+i \lambda \sum _{m\epsilon Z } \rho ( m) \Pi ( m )  \nonumber 
\end {eqnarray} 
\noindent 
Where the function $ \rho ( m ) $  extends  the function  $ \sigma ( m ) $ which vanishes at m=0 ,as. 
\begin {equation}       
\rho ( m ) =\left \{ \begin{array}{cc} 
\sigma ( m )  & \mbox {$ m \neq 0$ }\\ 
-iK & \mbox { $ m=0$ } 
\end{array} 
                \right. 
\label{eq:rho}       
\end {equation}       
\noindent       
Note that to get the second equation in  eq ~\ref{eq:tc} from  
the first  one we use the commutation relation $ [ W,\Pi ( m )       
]=m \Pi ( m ) $ one may extract from the defining equations of the involved operators.We thus recover the       
formula  of the twisted winding number $W ( \lambda ,\rho )$  we found  in the first approach.       
\section{Group theoretic approach }      
In this section we would like to show       
how enhancing the fundamental group $ \Pi (m) $ from discrete to continuous in        
a specific but not trivial way has the effect of introducing interactions        
among loops and precisely those worked out in the hamiltonian above.The only limitation of the approach       
is that it corresponds to the spectral function in its simplest form of a phase $ \rho ( m ) =( - ) ^m $.The       
reason of this limitation is obvious as only when  loops  $ \mid m > $ are assigned equal weights (       
couplings ) $ \rho ( m ) $ up to a phase  does one expect them to still form a group structure and therefore 
to extract       
some group properties.       
The starting point is to rewrite the fundamental group elements in a form        
familiar to continuous groups.       
\begin {equation}       
\Pi( m )  = e^{-iJm}      
\label{eq:pi}       
\end {equation}       
\noindent       
J is an operator which will become the generator of the resulting        
continuous group.        
The action of the generator J on the $ \mid \theta > $ states  defined earlier can be inferred from eq    
~\ref{eq:pi}       
\begin {equation}       
J\mid \theta \rangle =\theta \mid \theta \rangle      
\label{eq:j}       
\end {equation}       
\noindent       
In subsequent analysis  we need  however the representation of J on the        
 space of the $ \mid n > $  states.This may be obtained        
by inverting   eq  ~\ref{eq:pi} which is invertible      
 .To this end  a simple way to proceed is to  multiply both sides of the equation by the        
factor  $  \frac{( - ) ^m}{m } $    then sum over all values m $ \epsilon  Z /(0)$ .       
\begin{eqnarray}  
 \label{eq:inversion}       
 \sum _{m\epsilon Z/(0) } ( -  )^m \frac{ \Pi ( m )}{m}& =& \sum _{m\epsilon Z/(0) } ( -  )^m \frac{e^{-
im\theta }}{m}       
\nonumber \\       
                                                            &=&-2i \sum_{m\epsilon Z/(0)} ( -  )^m \frac{sin ( m 
\theta)}{m}\\  
&=&i\theta  \;\; -\pi \; \langle \;\theta \; \langle  \; \pi \nonumber \\      
&=&iJ \nonumber  
\end{eqnarray}        
\noindent         
In the above sequence , both sides  of the equations are acting on the states        
$ \mid \theta >$ which we do not write explicitly for clarity .Thus  we end up with the expression for J we 
look for.      
\begin {equation}       
 J=-i\sum _{m\epsilon Z/(0) } ( -  ) ^m \frac{ \Pi ( m ) }{m}     
\label {eq:J2}      
\end {equation}        
\noindent         
The second key point and the most important is to make the transition from        
discrete values to continuous in the parameter space of the group ,while        
keeping the generator of the group unchanged.This leads immediately to the        
SO(2) group .       
\begin {equation}       
\Pi ( \lambda ) = e^{-iJ\lambda}      
\label {eq:rotation}       
\end {equation}      
\noindent       
It is important to realise that the generator J in the above equation is not modified and keeps its defining       
form as a sum over the $\Pi (m ) $ .       
Note that from the fact that the variable        
$ \theta $  is continuous the resulting group does not have periodicity in the $\lambda $ variable .       
The next step is to see how the new operators $ \Pi( \lambda ) $ will act on the old states $ \mid n >$ and       
write   down the expression for these,then  work out the operator "Hamiltonian"  $ W_\lambda $ which       
generates these states as eigenstates.Denoting the new states as $ \mid n_\lambda >$ we have the set of      
defining equations , both for $ \mid n_\lambda \rangle $ and $ W_\lambda$.      
\begin {eqnarray}       
\mid n_\lambda \rangle &=& e^{-iJ \lambda } \mid n \rangle \nonumber \\      
W_\lambda \mid n _\lambda \rangle &=& n_\lambda \mid n_\lambda \rangle      
\label{eq:definingeq}        
\end {eqnarray}        
\noindent         
The above states can be worked out  as we know how  J acts on the $ \mid n > $ states        
. An immediate consequence is that the new states by their very        
definition "collapse" to old states of well defined winding numbers when        
the angle $ \lambda $  picks up integer values i.e.       
\begin {equation}       
\mid n_\lambda \rangle _{\lambda =m}=\mid n+m \rangle      
\label {eq:collapse}      
\end {equation}       
\noindent        
This property is useful in that it allows the extraction of the unknown        
eigenvalues $n_\lambda $ in eq ~\ref{eq:definingeq} .Expanding  $ n ( \lambda ) $ in powers of $       
\lambda $  the expansion  should saturates  at the first order in $ \lambda$ otherwise it will contradict       
equation ~\ref{eq:collapse} .We thus obtain .      
\begin {equation}       
n_\lambda =n+\lambda      
\label {eq:neweigenvalues}      
\end {equation}       
\noindent        
It remains now to work out the explicit form of the operator $ W_\lambda$.Rewrite the        
eigenvalue equation ~\ref{eq:definingeq} as       
\begin {eqnarray}       
W_\lambda e^{-iJ \lambda } \mid n \rangle &=& n_\lambda e^{-iJ \lambda } \mid n \rangle  \nonumber \\      
                                                                    &=& e^{-iJ \lambda } ( W + \lambda ) \mid n \rangle       
\label{eq:effective2}        
\end {eqnarray}        
\noindent         
In operatorial form this equation reads.      
\begin {equation}      
W_\lambda =e^{-iJ \lambda } W e^{i J \lambda } + \lambda       
\label {eq:transformation1}      
\end {equation}       
\noindent        
To make   loop self-interactions  more  explicit,we use the familiar formula.       
\begin {equation}       
e^{-iJ \lambda } W e^{i J \lambda }=W+i\lambda [ W,J ]      
\label {eq:transformation2}      
\end {equation}       
\noindent        
Together with the commutation relations we have between W and the $ \Pi ( m ) $ .After some algebra we       
get        
the desired result  that .       
\begin {equation}       
W_\lambda = W + \lambda \sum _{m\epsilon Z } ( - )^m \Pi ( m )       
\label {eq:endresult}      
\end {equation}       
\noindent        
This is the effective winding number operator to which the external field $ \beta $        
couples as in ~\ref{eq:effective} in which the  angle $ \lambda $   plays the role of the        
self coupling.       
        
\end{document}